# High-temperature threshold of damage of SiC by swift heavy ions

D.I. Zainutdinov[1,2*], V.A. Borodin [2], S. A. Gorbunov[1], N. Medvedev[3,4], R. A. Rymzhanov [5,6], M.V. Sorokin[2], R.A. Voronkov[1], A.E. Volkov[1,2]

[1]*P.N. Lebedev Physical Institute of the Russian Academy of Sciences, Leninskij pr., 53,119991 Moscow, Russia*

[2] *NRC Kurchatov Institute, 123182 Moscow, Russia*

[3] *Institute of Physics, Czech Academy of Sciences, Na Slovance 1999/2, 182 21 Prague 8, Czech Republic*

[4] *Institute of Plasma Physics, Czech Academy of Sciences, Za Slovankou 1782/3, 182 00 Prague 8, Czech Republic*

[5] *Flerov Laboratory of Nuclear Research, Joint Institute for Nuclear Research, 141980 Dubna, Russia*

[6] *Institute of Nuclear Physics,050032 Almaty, Kazakhstan*

## Abstract

At ambient conditions, SiC is known to be resistant to irradiation with swift heavy ions (SHI) decelerating in the electronic stopping regime. However, there is no experimental data on the SiC irradiation at elevated temperatures. To investigate this problem, we evaluate the stability of SiC to SHI impacts at high temperatures up to 2200 K. We apply the combination of the Monte-Carlo code TREKIS-3, describing excitation of the electronic and atomic systems using temperature-dependent scattering cross-sections, with molecular-dynamic modeling of the lattice response to the excitation. We demonstrate that increasing irradiation temperature increases the energy transferred to the atomic lattice from the excited electronic system. This material heating leads to formation of a stable nanometric damaged core along the trajectory of 710 MeV Bi ion when the irradiation temperature overcomes the threshold of ~1800 K. In this case, a chain of nanometric voids along the ion trajectory forms due to the mass transport from the track core by edge dislocations. Voids of larger sizes appear at higher irradiation temperatures. At lower irradiation temperatures, the damaged regions recrystallize completely within ~100 ps after the ion passage.

---

*Corresponding author: d.zaynutdinov@lebedev.ru

# I.    Introduction

Swift heavy ions (SHI), with masses $> 10m_p$ ($m_p$ is the proton mass) and energies 1-10 MeV/nucleon decelerate in solids in the electronic stopping regime [1–3]. Energy transfer from the excited electronic system of a target to the atomic lattice may cause formation of a nanometric region with a modified structure along the ion path, known as an SHI track. This damage generates significant interest in SHI irradiation in the communities of the nuclear, nano-, and bio-technologies, as well as medicine [4–7].

The high melting point of SiC (~3100 K) makes it a promising material for applications in modern aerospace and nuclear technologies [8–10]. The material is also resistant to the irradiation with SHIs, at least at room temperature [11]. However, to the best of our knowledge, no results of SHI irradiation of silicon carbide at elevated temperatures are available. It is impossible to say a priori what effect of the elevated irradiation temperature to expect: On the one hand, it may lead to the annealing of defects and make the material more resistant [12]; on the other hand, it cannot be excluded that the SiC is less resistant to swift heavy ions at high irradiation temperatures due to synergistic effects of the target temperature and the transient local temperature increase in SHI tracks [13].

To get a better understanding of the effects accompanying the passage of swift heavy ions through SiC at high temperatures, we have performed numerical simulations in the framework of the well-tested multiscale modeling approach. The approach couples the Monte-Caro (MC) code TREKIS-3 [14,15] that describes the short-term excitation of the electronic and ionic systems of the target (during ~100 fs) [3], with the molecular-dynamic (MD) code LAMMPS [16] for simulation of the subsequent atomic lattice response to the excitation afterward, as described in Section II [17–20].

Also, in Section II we analyze the choice of the MD potential appropriate for SHI-irradiation simulation. We study the applicability of Tersoff-ZBL with modified repulsive part [21] vs Vashishta [22] interatomic potential for the description of the silicon carbide lattice evolution in SHI tracks. We show that the Vashishta potential is better suited for this purpose.

Section III discusses the construction of temperature-dependent lattice scattering cross sections for TREKIS-3. We use the dynamic structure factor (DSF) formalism to evaluate these cross sections, depending on collective responses of target atoms to excitation. The temperature dependence of the atomic DSF can be restored from the molecular dynamic simulations tracing the atomic trajectories at different lattice temperatures [23–25].

Section IV demonstrates that silicon carbide, which is resistant to SHI irradiation at room temperature, can be damaged by individual swift heavy ions at irradiation temperatures above ~1800 K.

# II.    Model

The lifetime of the electronic excitation in a SHI track is less than 50-100 fs, which is comparable with the shortest collective dynamical modes (optical phonons) of the atomic ensemble [3]. This allows us to describe track formation using a multiscale model consisting of two blocks. The MC code TREKIS-3 [14,15] simulates excitation and relaxation of the target electronic system interacting with the target atomic system around the ion trajectory. The profile of the energy deposited into the atomic system by the time of the electron cooling provides the initial condition for the molecular dynamic modeling (LAMMPS code [26]) of the atomic lattice response to the excitation.

### a.    Monte-Carlo simulation

The MC code TREKIS-3 is based on the asymptotic trajectory method of the event-by-event simulation of the SHI passage through matter with the ionization of a target along the projectile trajectory. It includes the transport of excited electrons and valence holes, their scattering on the lattice (transferring kinetic energy to the atoms), and target electrons (with the generation of secondary holes and electrons). Auger decay of core holes accompanied by electron emission, radiative decay with photon emission, and further photon transport and absorption are also modelled [14,15].

The TREKIS-3 uses scattering cross sections taking into account the collective response of the electronic and ionic systems of a target to the excitations in the linear response (first-order Born) approximation via the dynamic structure factor (DSF) formalism [27,28]. Due to the fluctuation-dissipation theorem [29], the DSF of the equilibrium system can be expressed via the loss function - the imaginary part of the inverse complex dielectric function of the target (CDF or loss function):

$S(\omega, q) \sim Im[-\varepsilon^{-1}(\omega, q)]/\left(1 - e^{-\frac{\hbar\omega}{k_b T}}\right)$, where the temperature factor $\left(1 - e^{-\frac{\hbar\omega}{k_b T}}\right)^{-1}$ appears from the fundamental DSF asymmetry, $\hbar\omega$ and $\hbar q$ are the energy and momentum transferred from the projectile to the target, and $\hbar$ is the Planck's constant. In our work, we combine cross sections with DSF and loss function, taking advantage of each of these representations.

In insulators and semiconductors, the loss function has a bimodal form with the peaks separated in the transferred energy space and representing the responses of the atomic system ($\hbar\omega \lesssim 0.2$ eV) and the electronic one ($\hbar\omega \gtrsim E_{gap}$, where $E_{gap}$ is bandgap width of the material). It results in the separation of the cross section into two parts: the "elastic" scattering on the atomic system ($\sigma_{at}$, without excitation of electrons) and the "inelastic" one, exciting the electronic system ($\sigma_e$):

$$\frac{d^2\sigma}{d(\hbar q)d(\hbar\omega)} = \frac{d^2\sigma_{at}}{d(\hbar q)d(\hbar\omega)} + \frac{d^2\sigma_e}{d(\hbar q)d(\hbar\omega)}.$$

It is convenient to represent the inelastic scattering cross section through the electronic part of the loss function [3]:

$$\frac{d^2\sigma_e}{d(\hbar q)d(\hbar\omega)} = \frac{2\left[Z_{eff}^2(v)e\right]^2}{\hbar^2\pi v^2 n_e}\frac{1}{\hbar q}\left(1 - e^{-\frac{\hbar\omega}{k_b T}}\right)^{-1} Im\left[-\frac{1}{\varepsilon_e(\omega, q)}\right], \tag{1}$$

Here $v$ is the projectile velocity, $e$ is the electron charge, $n_e$ is the concentration of the scatterers (electrons), and $Im[-\varepsilon_e^{-1}(\omega, q)]$ is the electronic part of the loss function. The effective charge of the incident SHI, $Z_{eff}(v)$, is evaluated in the Barkas approximation [3,30],

$$Z_{eff}(v_{ion}) = Z_{ion}\left[1 - exp\left(-\frac{v_{ion}}{v_0}Z_{ion}^{-\frac{2}{3}}\right)\right], \tag{2}$$

where $Z_{ion}$ is the ion's atomic number, $v_0 = \frac{c}{125}$, where $c$ is the speed of light in a vacuum. The effective charges of electrons and valence holes are $Z_{eff} = 1$.

For the elastic scattering, one can also use Eq. (1) with the phonon part of the loss function ($Im\left[-\varepsilon_{ph}^{-1}(\omega, q)\right]$) and the concentration of atoms $n_{at}$ instead of $n_e$. However, in our work, it is important to consider the temperature dependence of the scattering cross section. For that reason, we use the DSF form of the cross section:

$$\frac{d^2\sigma_{at}}{d(\hbar q)d(\hbar\omega)} = \frac{q}{2\pi\hbar^4 v^2}\left|U_{p-at}(q)\right|^2 S_{at}(\omega, q). \tag{3}$$

Here $U_{p-at,}(q) = 4\pi e^2/(q^2 + (l_{scr})^{-2})$ is the interaction potential of the projectile (electron or valence-band hole) with an isolated target atom screened by valence electrons, with the screening parameter $l_{scr}$=0.14 nm (calculated with the methods from Ref. [25]) providing the agreement of the calculated cross section with that from the phonon part of the loss function (see below). The interaction of an incident SHI with the atomic system (nuclear stopping) was not taken into account.

Electron and valence hole scattering are not the only mechanisms of energy transfer to the atomic lattice of solids under extreme electronic excitation. Such excitation considerably changes the interatomic potential that accelerates the atoms finding new equilibrium positions, which increases the kinetic energy of the lattice ("nonthermal" heating) [31,32]. In covalent materials, this process is accompanied by the bandgap collapse at times comparable to or shorter than that of the cooling of the electronic system in tracks [3], allowing a description of the nonthermal lattice heating by conversion of the potential energy of electron-hole pairs into the kinetic energy of the lattice atoms at the time of the electronic system cooling down (~100 fs after the ion impact) [3,31,32].

The main result of the MC simulation is the radial distribution of the energy density transferred to the lattice around the ion trajectory through the two channels described above. TREKIS-3 also predicts the temporal evolution of this distribution, as well as the excited electron and valence hole density distributions during the relaxation time of the electronic excitation (~100 fs). All distributions are averaged over $10^3$ MC iterations for reliable statistics.

### b. Loss function from optical data at room temperature

As is conventional for modern transport MC models, the loss function is reconstructed from the experimental optical data following the Ritchie and Howie algorithm [33]. In the optical limit $q$=0, the total loss function $Im[-\varepsilon^{-1}(\omega, 0)]$ including phonon and electron parts can be restored from experimental data [14], using the absorption and refraction coefficients measured for low-energy photons (<50 eV [34]) and the photon absorption lengths $\lambda_{ph}$ for high-energy photons [35]. Then, the analytical form of this loss function can be constructed as a sum of the Drude-Lorentz oscillators [14]:

$$Im\left[-\frac{1}{\varepsilon_{ph,e}(\omega, q)}\right] = \sum_i \frac{A_i \gamma_i \hbar \omega}{\left(\hbar^2 \omega^2 - \left(E_{0i} + \frac{\hbar^2 q^2}{2M}\right)^2\right)^2 + (\gamma_i \hbar \omega)^2} \ , \qquad (4)$$

where $E_{0i}$ is the energy of the $i$-th collective oscillation mode of the target, $A_i$ is its intensity, $\gamma_i$ is the attenuation coefficient that determines the lifetime $\tau_i \sim \frac{1}{\gamma_i}$ of the $i$-th mode, and $M$ is the scatterer mass (electron or atom). The free particle dispersion law is used in Eq.(5) to extend the loss function to the entire plane ($\omega, q>0$) [15].

**Table 1** presents the fitting coefficients for SiC, and the values of $kk$ and $f$ sums [14], indicating a good quality of the fitting. Eq. (5) corresponding to the total $kk$-sum rule gives 1.04568 (4.568% deviation from 1), and Eq. (6) corresponding to the total $f$-sum rule for electrons produces 20.158 (only 0.79% deviation from the total number of electrons 20).

1. The $kk$ – sum rule (or $ps$ – sum rule) for the total loss function states that:

$$\frac{2}{\pi}\int_{0}^{+\infty} Im[\varepsilon^{-1}(\omega, q = 0)]\frac{d(\hbar\omega)}{\hbar\omega} \rightarrow 1 \tag{5}$$

2. The $f$–sum rule is:

$$\frac{2}{\pi(\hbar^2\Omega_p^2)}\int_{0}^{+\infty} Im[\varepsilon^{-1}(\omega, q = 0)](\hbar\omega)d(\hbar\omega) = N_{eff} \tag{6}$$

For the electronic part of the loss function, $N_{eff}$ must be equal to the number of valence or core-shell electrons in the target molecule and $\Omega_p^2 = 4\pi n_m e^2/m_e \approx 8\ eV$ is the plasma frequency, where $n_m$ is target molecule density, $m_e$ is the free electron mass. For the phonon part of the loss function, $N_{eff}$ must be equal to unity, and $\Omega_p^2 = 4\pi e^2 \sum_\alpha n_\alpha/M_\alpha \approx 0.06\ eV$, where the sum runs over all types of atoms in the target, and $n_\alpha$ and $M_\alpha$ are the density and mass of the α-th type of atoms, respectively.

*Table 1 The coefficients of the loss function in the form of the sum of Drude oscillators (Eq. (4)) providing a good agreement with the experimental optical data for 6H-SiC [34,35].*

|  | $A_i(eV^2)$ | $\gamma_i(eV)$ | $E_{0i}(eV)$ | $kk$ – sum | $f$ – sum $(N_{eff})$ |
|---|---|---|---|---|---|
| Phonons: | $3\times 10^{-3}$ | $2\times 10^{-3}$ | 0.121 | 0.20490 | 0.73(1) |
| Valence band: | 90 | 80 | 29.39 | 0.82435 | 8.054(8) |
|  | -7.565 | 12.327 | 8.075 |  |  |
|  | 454.5 | 7 | 23.347 |  |  |
| L-Si | 580.05 | 120 | 145.009 | $1.543\times 10^{-2}$ | 8.014(8) |
| K-C | 185 | 100 | 303 | $0.99\times 10^{-3}$ | 2.072(2) |
| K-Si | 175 | 900 | 2000.044 | $2\times 10^{-5}$ | 2.018(2) |
| Result: |  |  |  | 1.04568 | 20.158 |

The experimental CDF for SiC and its analytical approximation are shown in **Figure 1**.

Because the optical experiments were overwhelmingly performed at room temperature, the reconstructed loss function (i.e., $A_i$, $E_{0i}$, $\gamma_i$ coefficients) is valid at this temperature. The available high-temperature optical experimental data for SiC [36,37] consider only small energy and temperature ranges (up to 773 K) and do not provide complete information about the loss function temperature dependence. Therefore, we cannot directly use these data in our work. Thus, to solve the temperature dependence problem, we use the approximations discussed below.

First, we assume that the temperature dependence of the electronic part of the loss function can be neglected [3] at target temperatures below the melting point ($T_{melt} \sim 3100$ K [38]). At the same time, we take into account the temperature dependence of the band gap width, which shrinks from 2.77 eV at 300 K to 2.5 eV at high temperatures [39].

Second, having in mind that for the elastic scattering, the DSF is the Fourier transform of the spatial-temporal pair correlation function, it can be calculated at arbitrary temperature based on molecular dynamics tracing of classical atomic trajectories, as discussed below in Section III.

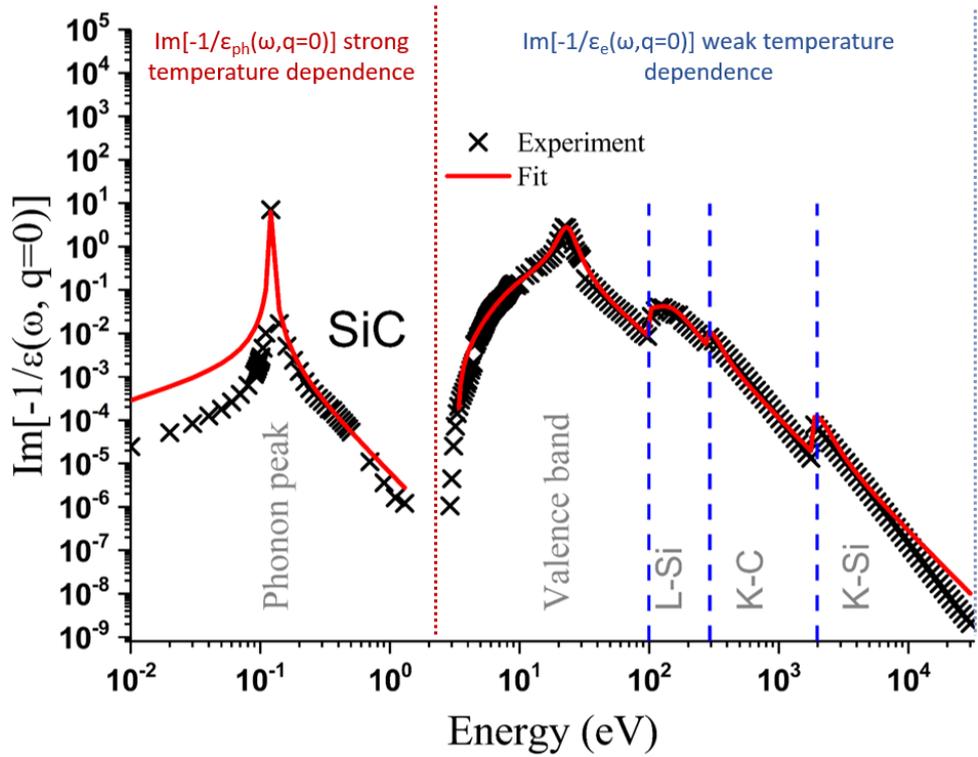

**Figure 1.** *The experimental* [34,35] *and reconstructed loss functions of 6H-SiC.*

### c. MD model

For simulation of the SHI track formation, we use the rectangular parallelepiped simulation cell containing 6H-SiC crystal with the crystalline axes [10$\bar{1}$0], [$\bar{1}$2$\bar{1}$0] and [0001] along the simulation cell sides with periodic boundary conditions. The supercell size is $37 \times 38.5 \times 12.1$ nm$^3$ containing 1,658,880 atoms.

Prior to the track simulations, the bulk supercell was equilibrated at the desired temperature using a two-step procedure [40]: (i) equilibration of the supercell size in the NPT ensemble at atmospheric pressure for over 30,000 timesteps; and (ii) further equilibration in the NVE ensemble for over 10,000 timesteps. The integration step of 0.2 fs was used in all steps, as suggested in Ref. [40].

The SHI track axis was assumed to be aligned along the <0001> axis of silicon carbide crystal. The passage of SHI itself was not simulated explicitly in MD. Instead, it was assumed that at the time scales of MD simulation, the main effect of the SHI passage was the instantaneous heating of the ions in the vicinity of the track axis. In order to assign the initial atomic velocities for the subsequent structural evolution, the supercell volume around the ion trajectory was divided into cylindrical layers centered on the track axis. The atoms within each layer were assigned randomly directed momenta with the absolute value sampled from the Gaussian distribution with the average kinetic energy equal to the radial energy transferred to the atomic system predicted by TREKIS-3 for each particular layer [17].

The evolution of the atomic structure after the passage of SHI was simulated using an NVE ensemble with the integration step of 0.2 fs until the track core cooled down to the irradiation temperature (~ 200-300 ps). During the simulation, the Berendsen thermostat [41] with the damping time of 20 fs was applied in the layer of 5 Å thickness at the cell boundaries parallel to the ion trajectory. Since this layer is far away from the cell center, it has little effect on the atomic excitation evolution around the SHI trajectory.

### d. Selection of interatomic potential

Ultrafast lattice heating generates high pressures in the proximity of the SHI trajectory, which means that the applied interatomic potential should be able to adequately describe short-range interatomic repulsion. Among the available potentials for SiC [21,22,42–46], this requirement is satisfied only by Tersoff-ZBL [21] and Vashishta [22] potentials developed for the radiation physics [47] and high-pressure problems [48], respectively.

To validate the applicability of these potentials for the simulation of SHI tracks, we can use the experimental fact that no SHI tracks form in SiC crystals irradiated at room temperature with 710 MeV Bi ions [49]. The results of TREKIS+MD simulations, corresponding to these experimental conditions (see **Figure 2(a,b)**), indicate that only Vashishta potential predicts correctly the lack of noticeable crystal damage formed. In contrast, the Tersoff-ZBL potential results in a strongly damaged region of ~6 nm size around the ion trajectory after the track cooling down. Both potentials predict the same size of the transiently disordered region of ~6 nm, but only the Vashishta potential describes its subsequent recovery taking place within the first 100 ps after the ion passage. Thus, the Vashishta interatomic potential is used in the subsequent simulations.

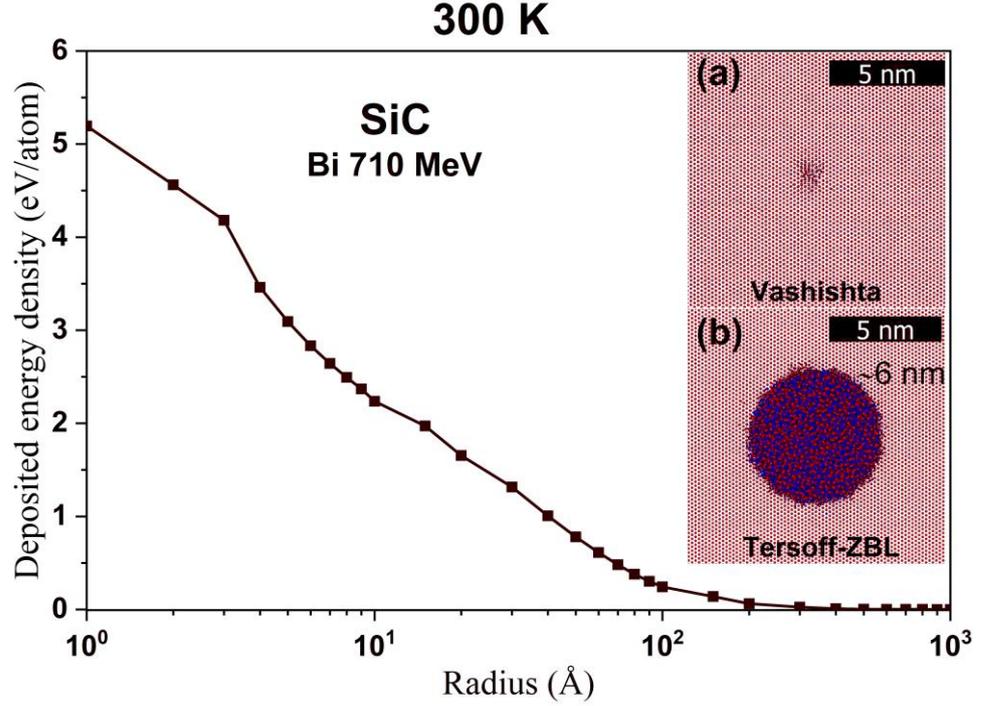

**Figure 2.** *Radial distribution of the total energy density transferred to the SiC lattice 100 fs after the 710 MeV Bi ion impact at 300 K. Subsequent structural changes in the SiC lattice after 100 ps obtained with (a) Vashishta potential [22], and (b) Tersoff-ZBL potential [21], are shown in the insets. The images are shown in the direction of ion incidence.*

## III.    Temperature-dependent atomic dynamic structure factor of SiC

The atomic DSF is the Fourier transform of the spatiotemporal pair correlation function of the atomic ensemble $G_{at}(t, \vec{r})$ of the target:

$$S_{at}(\omega, \vec{q}) = \frac{1}{2\pi} \int e^{i(\vec{q}\vec{r} - \omega t)} G_{at}(t, \vec{r}) d\vec{r} dt \,, \tag{7}$$

$$G_{at}(t, \vec{r}) = N_{at}^{-1} \langle \sum_{i,j=1}^{N_{at}} \int d\vec{r}' \delta(\vec{r} + \widehat{\vec{r}}_i(0) - \vec{r}') \delta(\vec{r}' - \widehat{\vec{r}}_j(t)) \rangle \,. \tag{8}$$

Here $N_{at}$ is the number of atoms in the system, $\{\widehat{\vec{r}}_i(t)\}$ are the atomic coordinate operators in the Heisenberg representation $\widehat{\vec{r}}_i(t) = \exp\left(\frac{i\widehat{H}t}{\hbar}\right)\widehat{\vec{r}}_i \exp\left(-\frac{i\widehat{H}t}{\hbar}\right)$ with the system Hamiltonian $\widehat{H}$. The angular brackets, $\langle ... \rangle = \sum_i p_i \langle i| ... |i\rangle$, represent the quantum and statistical averaging over the atomic ensemble, where $p_i$ is the statistical weight of the $i$-th ensemble quantum-mechanical state [27].

In the classical limit, the operators $\{\widehat{\vec{r}_i}(t)\}$ reduce to atomic trajectories $\{\vec{r}_i(t)\}$ which can be computed in MD simulations at arbitrary target temperatures [23–25]. Statistical averaging reduces to averaging over time.

For a multi-component system, the classical pair correlation function has the following form [50]:

$$G_{at}^{cl}(t,\vec{r}) = N_{at}^{-1} \sum_{\alpha=1}^{A} \sum_{\beta=1}^{A} Z_\alpha Z_\beta G_{\alpha\beta}^{cl}(\vec{r},t). \tag{9}$$

where $A$ is the number of different atomic species in the system ($A$=2 for SiC). $Z_\alpha$ and $Z_\beta$ are the model charges of the lattice ions in MD potential ($Z_{Si} = 1.201$ and $Z_C = -1.201$ for SiC [22]), while $G_{\alpha\beta}^{cl}(t,\vec{r})$ is the classical pair correlation function for $\alpha$ and $\beta$ atomic species that has the following form:

$$G_{\alpha\beta}^{cl}(t,\vec{r}) = \langle \sum_{i=1}^{N_\alpha} \sum_{j=1}^{N_\beta} \delta\left(\vec{r} + \vec{r}_i^\alpha(0) - \vec{r}_j^\beta(t)\right)\rangle. \tag{10}$$

We calculated the classical trajectories in SiC for Eq. (10) with MD simulations [25], constructing a database at various temperatures for the pre-relaxed SiC supercell with $N_{at} = 2880$ in the NVE ensemble [22]. The supercell equilibrium size changes weakly with the temperature and was $31 \times 32 \times 30$ Å$^3$ at $T$=300 K. Following the ergodic hypothesis, in order to obtain a statistically meaningful result, averaging over time in Eq. (10) was performed using the algorithm described in Ref. [24]. The calculated classical DSFs were averaged along the directions of $\vec{q}$; a possible influence of lattice anisotropy is beyond the scope of this work.

Due to the non-commutativity of the coordinate operators of atoms in Eq. (8) at different time instants, the isotropic quantum atomic DSF is a frequency-asymmetric function. In the thermalized atomic ensemble, this asymmetry has the form [51]:

$$S_{at}(\omega,q) = e^{\frac{\hbar\omega}{T}} S_{at}(-\omega,q). \tag{11}$$

In order to restore the fundamental asymmetry of the DSF calculated from the classical atomic trajectories, quantum correction factors are used [52–60]. The most commonly used correction factors suggested in the literature are listed in **Table 2**.

**Table 2** *Correction factors restoring the quantum asymmetry of the classical DSF [52–60].*

| | |
|---|---|
| $$S_{at}(\omega, q) = 2\left(1 + \exp\left(-\frac{\hbar\omega}{k_b T}\right)\right)^{-1} S_{at}^{cl}(\omega, q)$$ | Standard [54,56–58] |
| $$S_{at}(\omega, q) = \frac{\hbar\omega}{k_b T}\left(1 - \exp\left(-\frac{\hbar\omega}{k_b T}\right)\right)^{-1} S_{at}^{cl}(\omega, q)$$ | Harmonic [54,59,60] |
| $$S_{at}(\omega, q) = \exp\left(\frac{\hbar\omega}{2k_b T}\right) S_{at}^{cl}(\omega, q)$$ | Schofield [52] |
| $$S_{at}(\omega, q) = e^{\frac{\hbar\omega}{2k_b T}} \int_{-\infty}^{+\infty} e^{-i\omega t} \int_{-\infty}^{+\infty} e^{i\omega'\sqrt{t^2 + \left(\frac{\hbar}{2k_b T}\right)^2}} S_{at}^{cl}(\omega', q) d\omega' \frac{dt}{2\pi}$$ | Egelstaff [53] |

However, among these possibilities, only the *«Harmonic»* correction factor satisfies the sum rule (see **Figure 3(a)**) that the quantum DSF must obey based on its definition by Eq. (7-8), namely [61]:

$$\int_{-\infty}^{+\infty} \omega \cdot S_{at}(\omega, q)\, d\omega = \frac{\hbar q^2}{2} \sum_{\alpha=1}^{A} Z_\alpha^2 \frac{x_\alpha}{m_\alpha}. \tag{12}$$

where $m_\alpha$, $x_\alpha$ are the mass and atomic fraction of the $\alpha$-type atoms in the simulation cell ($x_{Si} = x_C = 1/2$).

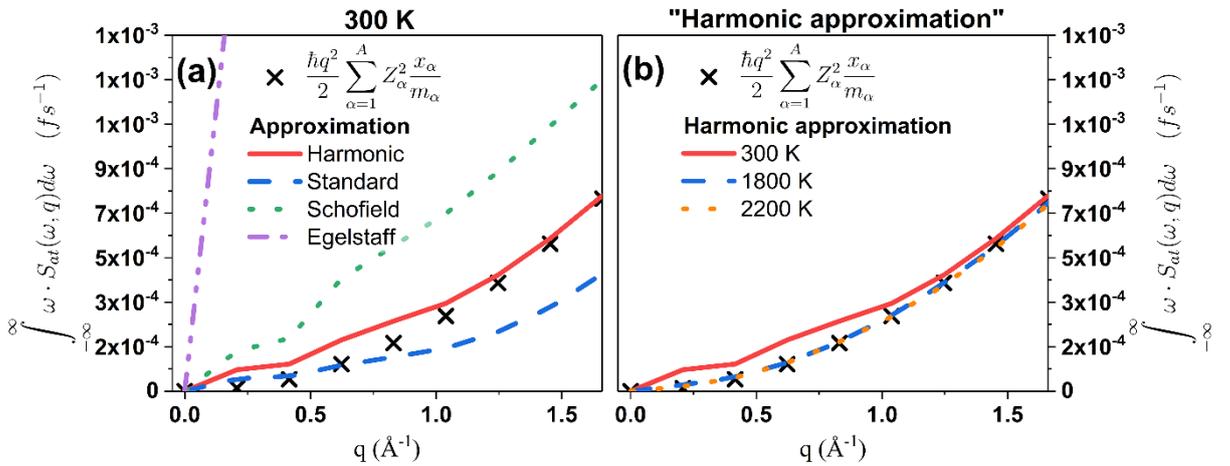

**Figure 3.** *(a) Comparison of the frequency integral $\omega \cdot S(\omega, q)$, for $S(\omega, q)$ of SiC calculated with different correction factors from **Table 2**, with the sum rule; and (b) the same comparison at different temperatures for the «Harmonic» correction factor.*

Since Eq. (12) is temperature independent, we checked the sum rule with the *«Harmonic»* factor over a wide temperature range (see **Figure 3(b)**), finding a satisfactory agreement. Small deviations at short *q* at room temperature are caused by artifacts due to the periodic boundary conditions on the scale of the simulation cell size and can be neglected [62]. Thus, we use the atomic DSF with the *«Harmonic»* correction factor in this paper:

$$S_{at}(\omega, q) = \frac{\hbar\omega/k_b T}{1 - exp\left(-\frac{\hbar\omega}{k_b T}\right)} S_{at}^{cl}(\omega, q). \tag{13}$$

**Figure 4** and **Figure 5** demonstrate that the DSF peaks increase in height and width and shift to lower frequencies with temperature increasing from 300 K to 2200 K. A similar peak behavior was observed in the Raman spectroscopy experimental data [63,64].

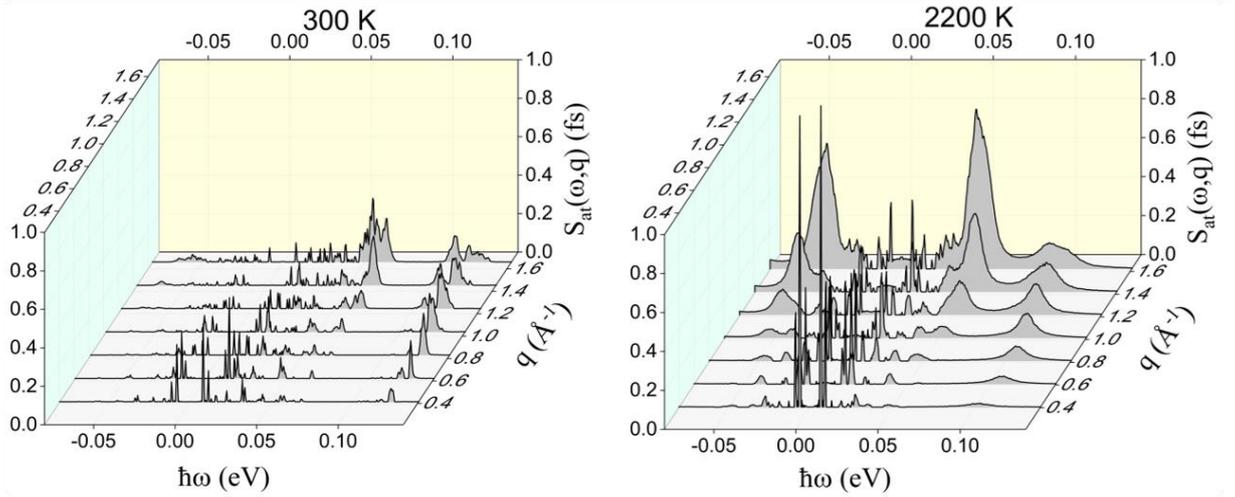

**Figure 4.** *DSF of the SiC atomic system with the «Harmonic» correction factor at different lattice temperatures (300 K and 2200 K).*

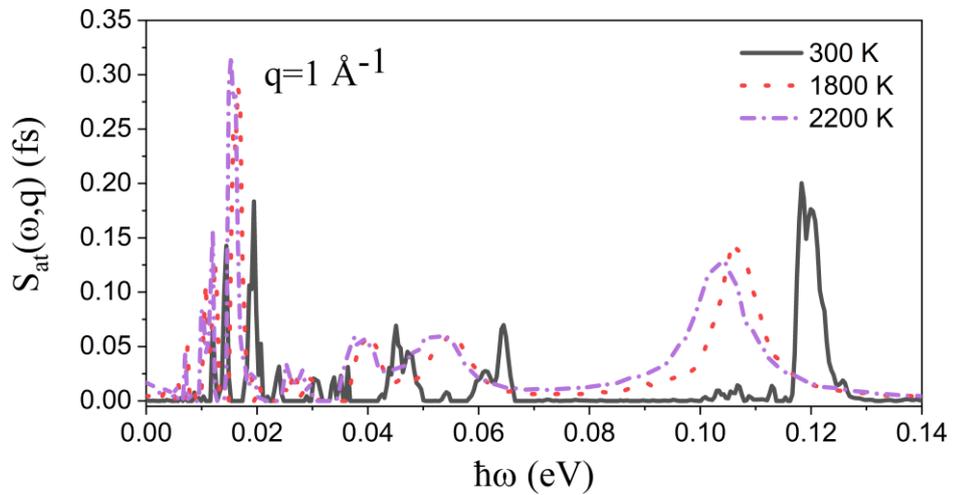

**Figure 5.** *Positive frequency part of the calculated DSF of the SiC atomic system with the «Harmonic» correction factor at different temperatures.*

**Figure 6(a,b)** shows the mean free path lengths of an electron and valence-band hole elastic scattering on the atomic lattice $\lambda_{at}^{el,h} = 1/(n_{at}\sigma_{at}^{el,h})$ obtained from Eq.(3) [15,65], applying the atomic DSFs with the «*Harmonic*» correction factor calculated at different temperatures. This mean free path decreases with temperature, which can be associated with an increase in the number of phonons enhancing the scattering. At room temperature, the mean free paths calculated with the atomic loss function (CDF) and those calculated with DSF (with the «*Harmonic*» correction factor) coincide at energies above 0.2 eV, validating our calculations. The difference at the lower energies is due to the existence of acoustic atomic modes in the calculated DSF, which are absent in the loss function restored from the optical data where only optical phonons contribute.

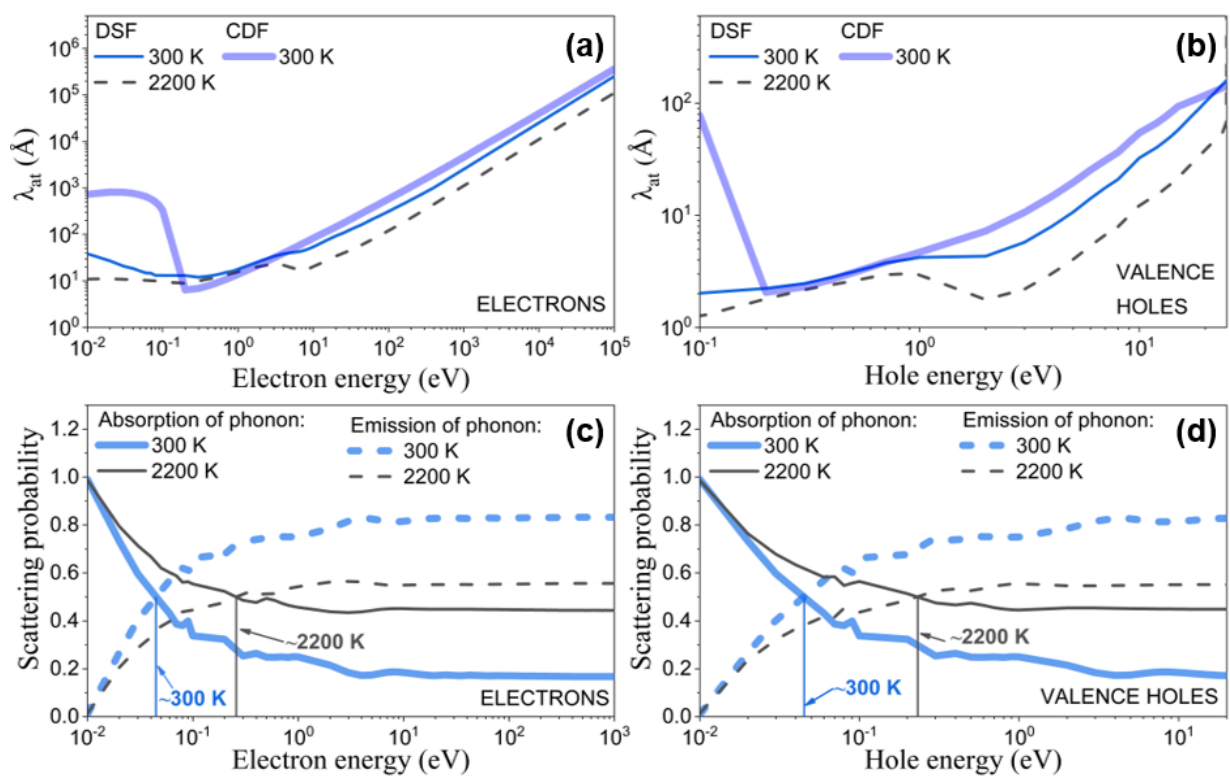

**Figure 6.** *Elastic mean free paths of electrons (top left panel) and valence holes (top right panel). Scattering probabilities of electrons (bottom left panel) and valence holes (bottom right panel) on the atomic lattice with phonon absorption and emission in SiC.*

Each scattering event on the lattice leads to phonon excitation (emission) or absorption. The first process heats the lattice while the second one cools it down. **Figure 6(c,d)** show the probabilities of phonon emission and absorption, which are defined as the ratios of the cross sections of these processes to the total elastic scattering cross section. The phonon absorption is dominant over the emission for particles with energies below the average kinetic energy of the target atoms defined by the irradiation temperature. As the lattice temperature increases, the energy range of such electrons also increases. In contrast, particles with higher energies preferentially heat the lattice by phonon emission during the scattering.

Additionally, the DSF becomes more symmetric (see Eq. (11)) when the temperature increases, which leads to an increase in the fraction of collisions with phonon absorption (see **Figure 6(c,d)**). Alongside the increase in the total number of collisions, this determines the profile of the energy transferred to the lattice due to the scatterings, which is discussed below.

# IV.    Results and Discussion

## a.    Excitation of SiC in SHI tracks at 300 K vs 2200 K

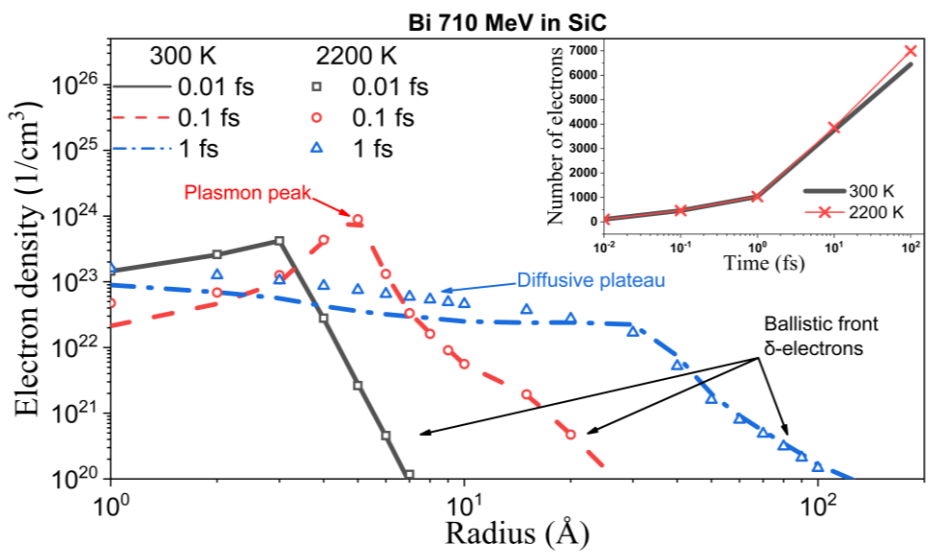

**Figure 7.** *Radial distribution of excited electrons around the Bi 710 MeV ion trajectory in SiC at different times; the total number of electrons shown in the inset.*

The effects of temperature are most pronounced at 2200 K, which is the highest temperature considered in this paper. **Figure 7** compares the evolution of the excited electron density until 1 fs after the ion passage around the trajectory of 710 MeV Bi ion (with the electronic stopping $\frac{dE}{dx} \approx$ 35 keV/nm) in SiC crystals irradiated at 300 K and 2200 K. The curves for 300 K and 2200 K are very similar up to this time. Up to 0.01 fs, the ballistic spreading of delta electrons formed by the ion passing through the target is pronounced [14,15]. Further, up to 0.1 fs, a new front is formed behind the ballistic one, associated with the plasmon excitations and decays [14,15]. We conclude that the irradiation temperature has little effect on the fast electron transport at these initial times.

That is because the excited electrons preferentially scatter inelastically [3], and this channel is very weakly temperature dependent since it is accompanied by high transferred energies to overcome the SiC band gap $\hbar\omega > E_{gap} \gg k_b T_{irr}$, where $T_{irr}$ is the irradiation temperature. Presenting the temporal dependence of the number of excited electrons that appeared in the proximity of the ion trajectory, the inset in **Figure 7** confirms a little temperature effect on the ionization channel.

The share of "slow" electrons moving behind the fast $\delta$-electrons increases with time. Starting from ~1 fs, their diffusive transport due to the interaction with the lattice leads to the formation of a plateau-like distribution presented in **Figure 7** [15]. At the same time, the high density of electrons in the central region of the track at high irradiation temperature is associated with a deceleration of their diffusion due to the increase in frequency of their collisions with the lattice. Similar trends are also observed in the transport of valence holes.

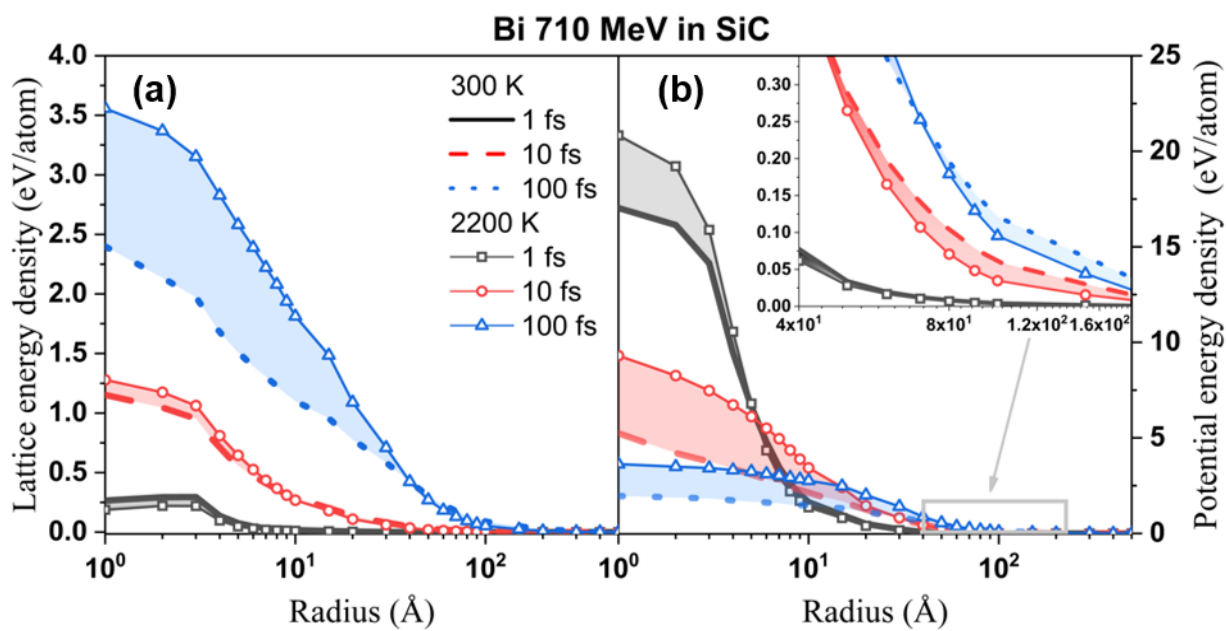

**Figure 8.** *Radial dependence of a) the density of energy transferred to the SiC atomic lattice by scattering electrons and valence holes, and b) the potential energy density of electron-hole pairs in the track of 710 MeV Bi ion at different times. The shaded areas highlight the difference between the profiles at different irradiation temperatures at the same time instants.*

The scattering of electrons and valence holes on phonons leads to the lattice heating. As a result, the increase in the irradiation temperature enhances the scattering of generated electrons and valence holes increasing the energy transfer rate and, hence, the density of the transferred energy in the lattice around the ion trajectory (see **Figure 8(a)**).

In addition, the nonthermal heating of the lattice also changes with the irradiation temperature, since it is associated with the electron-hole pairs density by the time of cooling down of the electron system in the track (∼100 fs). The increase in the potential energy density of electron-hole pairs located closer to the track core with temperature leads to an increase in lattice heating through the nonthermal channel, as can be seen in **Figure 8(b)**. On the other hand, the inset in **Figure 8(b)** shows that the electron-hole pair density decreases at a distance >40 Å. This leads to an increase in the temperature gradient directed toward the track core.

**Figure 9** presents the difference between the radial distributions of the energy density transferred to the atomic lattice of SiC by the time of 100 fs at different irradiation temperatures. We note that the room temperature profile calculated with the DSF-based elastic scattering cross-section coincides with that previously obtained with the loss function restored from the optical data. Thus, the results of the calculations with the temperature-dependent atomic DSF agree with those made with the well-validated approach [3] and the experimental data [49]. The figure shows the increase in the density of the energy deposited in the atomic lattice with the increase of the irradiation temperature.

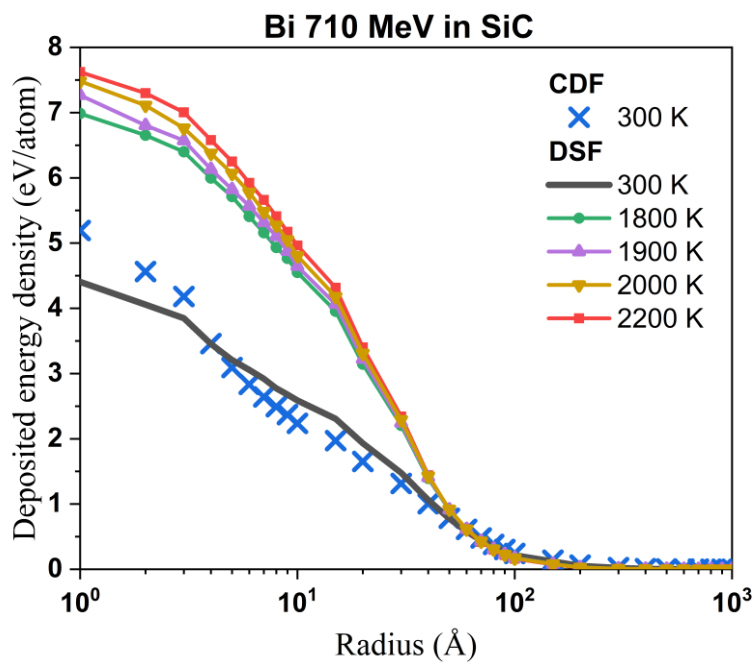

**Figure 9.** *The radial distributions of the total energy density deposited into the SiC lattice at different irradiation temperatures around the trajectory of 710 MeV Bi ion at 100 fs after the ion passage, obtained using the optical data based lattice cross sections at 300 K (CDF) and those using the MD calculated atomic DSF with the «Harmonic» correction factor at 300 K, 1800 K, 1900 K, 2000 K, 2200 K.*

### b. Structural changes in SiC irradiated at 300 K and 2200 K

**Figure 10** illustrates the MD-simulated temporal evolution of SiC atomic lattice after the passage of a 710 MeV Bi ion at room temperature. After the recrystallization that took ~30 ps, only a few point defects remained in the transiently damaged region of ~7 nm diameter. The recrystallization front moves from the periphery to the center of the track, as was also observed in other materials (e.g. MgO) in Ref. [66]. It means that the recrystallization is the dominant mechanism of the SiC resistance to SHI irradiation at room temperature. This fast recrystallization can be explained by the simple crystalline structure of 6H-SiC [66].

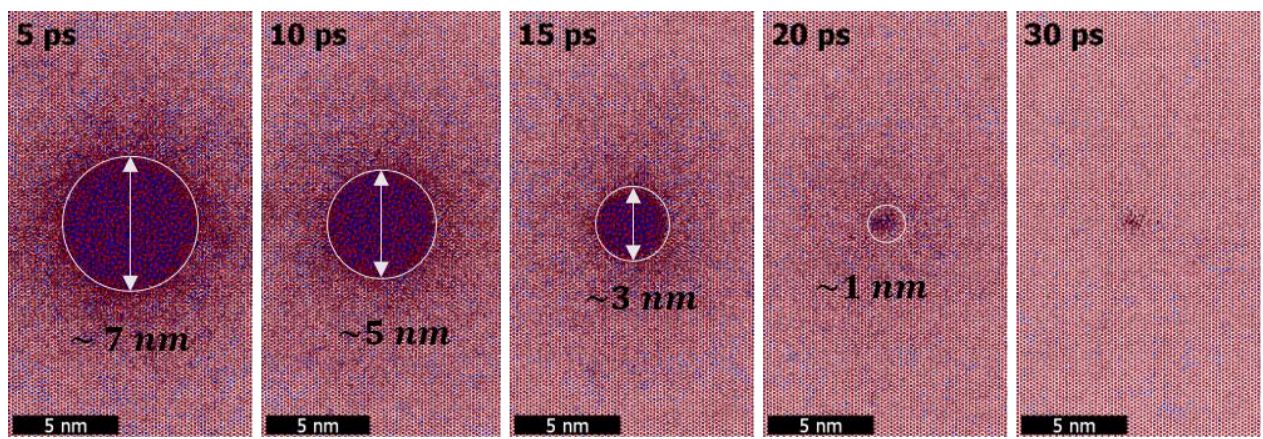

**Figure 10**. *Time evolution of atomic snapshots in SiC irradiated with 710 MeV Bi ion at 300 K.*

**Figure 11** shows that irradiation of silicon carbide at 2200 K leads to the formation of a primary damaged region of ~13.4 nm in diameter during the first 0.5 ps after the ion passage. In the next 1.5 ps, this region grows due to the surrounding lattice melting. At the same time, it reshapes to a regular hexagonal cylinder, which is compatible with the hexagonal symmetry of the (001) crystallographic planes of SiC.

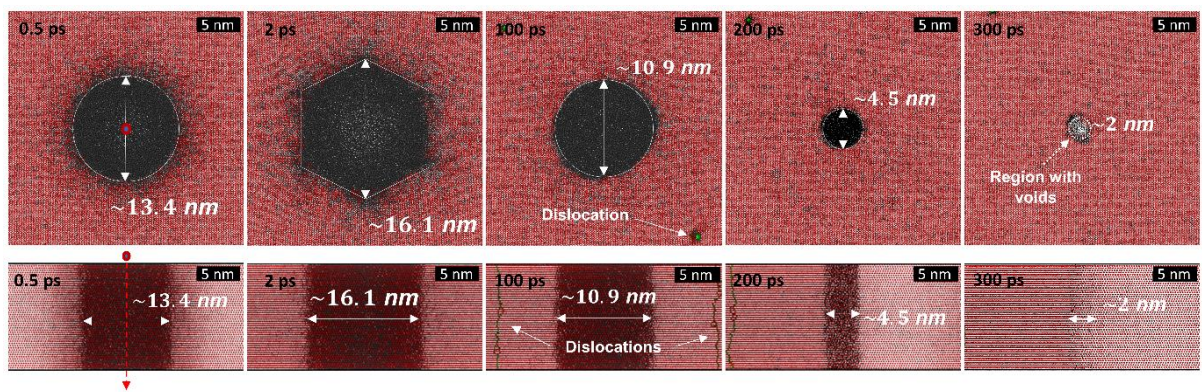

**Figure 11.** *Snapshots of the SiC carbon sublattice with dislocation lines colored and black defective region, taken at different times after 710 MeV Bi ion passage at 2200 K irradiation temperature.*

The density of amorphous silicon carbide at ambient conditions is 1.89 g/cm$^3$ [67], which is smaller than the crystalline one of 3.21 g/cm$^3$ [68]. Due to this difference, the formation of a disordered state in the track core creates a significant pressure of $P \approx 6.3$ GPa in the vicinity of the ion trajectory at the time of ~2 ps. This leads to the emission of dislocations from the hexagonal edges of the melted region (see **Figure 11**). Recrystallization starts at about the same time; by 200 ps after irradiation, small cavities are formed in the track core (see **Figure 12**). Over the next 100 ps, these cavities grow and coalesce into chains.

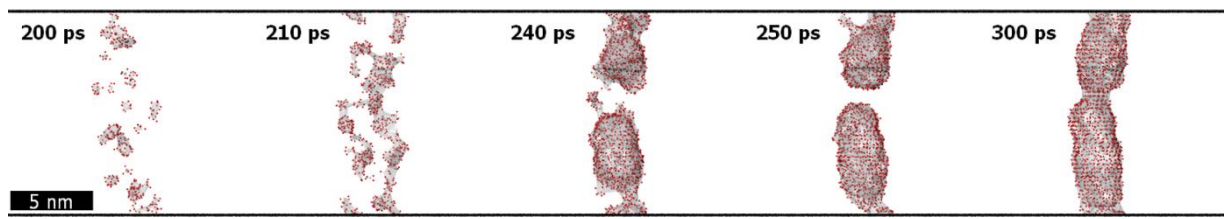

**Figure 12.** *Cavity surfaces in Bi 710 MeV ion track in SiC at 2200 K irradiation temperature at different times. Surface atoms are marked in red.*

As a result, an irregularly shaped cavity remains in the center of the track after the crystal cools down to the ambient temperature of 2200 K. Multiple point defects and defect clusters in both sublattices surround this region. The Wigner-Seitz defect analysis indicates that more defects are created in the lighter carbon sublattice. **Figure 13** presents the density profile and the track slice by the moment of 300 ps after the impact of 710 MeV Bi ion on SiC at 2200 K irradiation temperature.

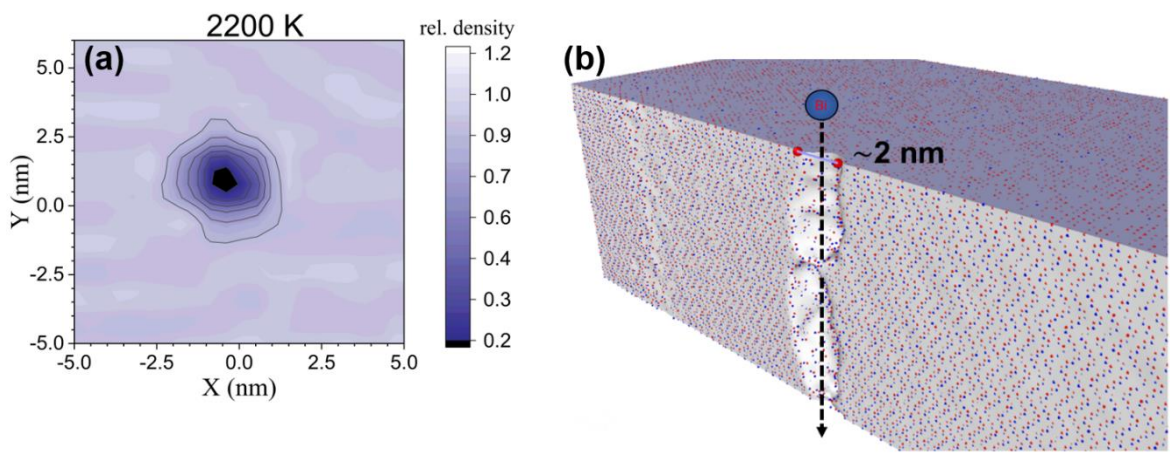

**Figure 13.** *(a) Relative density distribution in SiC irradiated at 2200K with Bi 710 MeV at 300 ps after the ion impact; (b) its longitudinal section.*

The cavity formation in the track core is the direct consequence of the mass transport away from the melted track region by edge dislocation emission. An example of the extra plane of such a dislocation is shown in **Figure 14**. A more detailed discussion on the dislocation formation and kinetics will be presented in the next subsection.

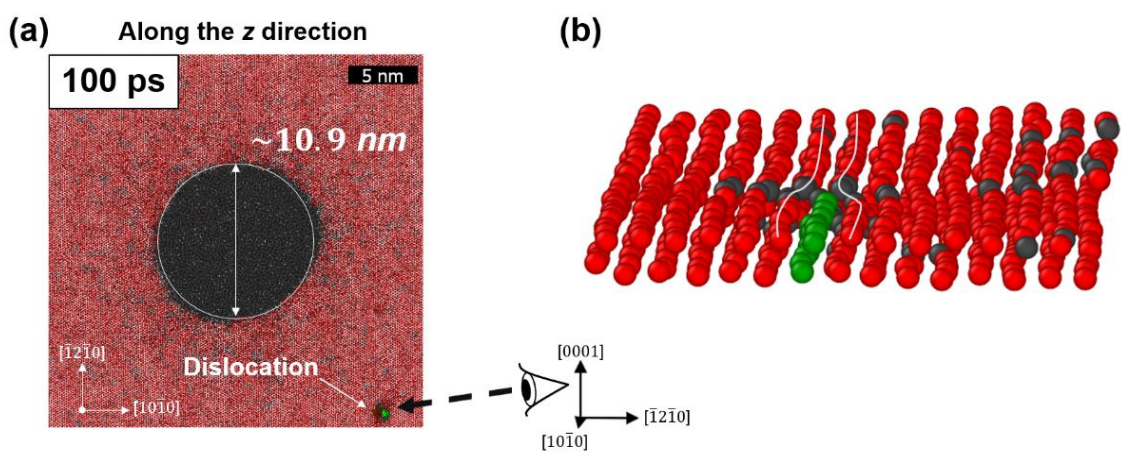

**Figure 14.** *(a) A snapshot of the SiC carbon sublattice in the direction of ion incidence at 100 ps after 710 MeV Bi impact at 2200 K with colored dislocation line and black defective region (b) A zoom at the edge dislocation extraplane (colored green) in this lattice surrounded by black interstitial atoms.*

### c. The threshold temperature of SiC damage

**Figure 15** shows the dependence of the relaxed diameter of the track of 710 MeV Bi ion in SiC on the irradiation temperature. To determine this dependence, we performed a set of calculations at 1800 K, 1900 K, 2000 K, and 2200 K irradiation temperatures. In all simulations, the track region passed through all the above-described stages. The calculations determined ~1800 K (1850±50 K) as the threshold temperature for the formation of residual damage regions in the SHI tracks in SiC.

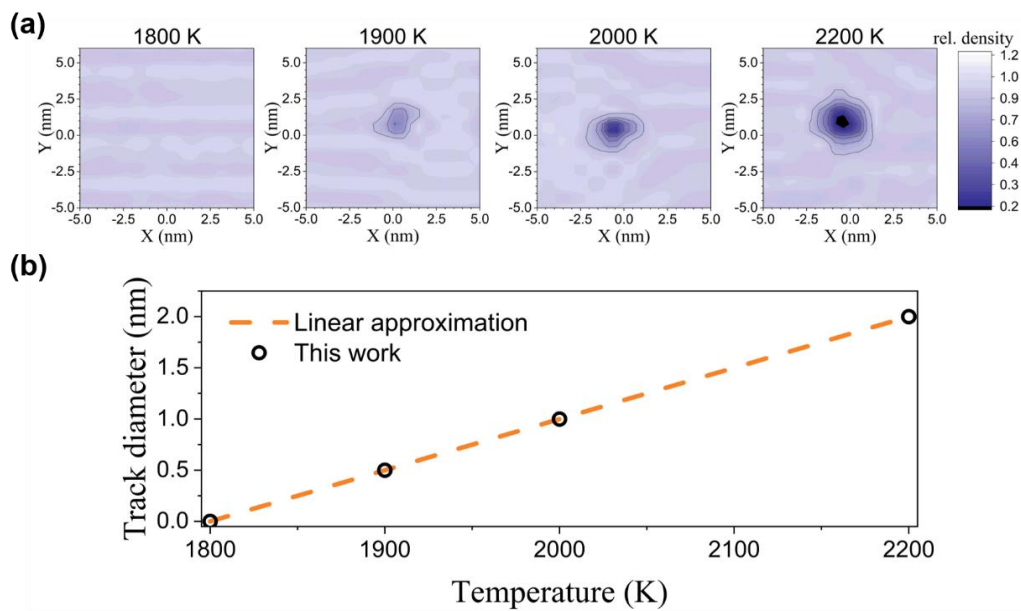

**Figure 15.** *(a) Relaxed spatial distributions of SiC relative density after irradiation with 710 MeV Bi ion at different temperatures; (b) Temperature dependence of the track diameter in SiC after 710 MeV Bi impact.*

The main reason for the incomplete recovery of the track region seems to be the transfer of material from the molten region in the track core to the bulk. The thermal expansion of the melt causes the local pressure to increase. This increase is more pronounced at the higher target temperatures because, as suggested by the MC estimates, the energy density deposited into the SiC lattice by an SHI increases with the target temperature increase. **Figure 16** shows the pressure distribution in the supercell at different temperatures at ~10 ps after the SHI passage, which is associated with the stress concentration at the hexagonal edges of the melt cylinder.

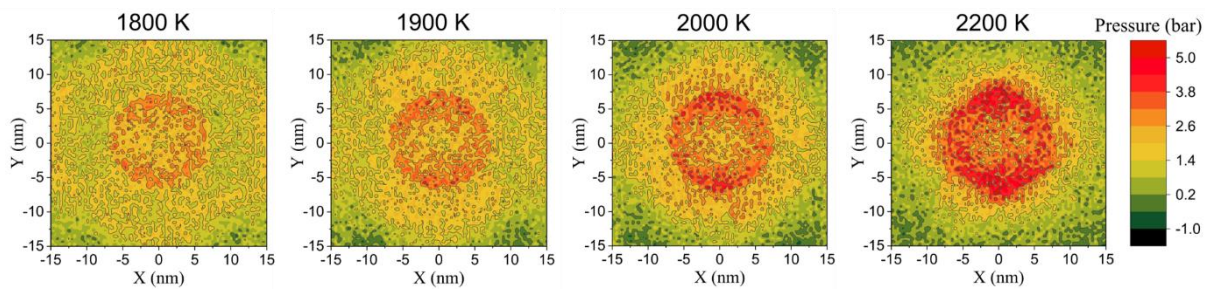

**Figure 16.** *Pressure distribution in the SiC supercell 10 ps after impacts of 710 MeV Bi at different temperatures.*

The dislocations emitted from the edges of the melt region are shown in **Figure 17**. These are edge dislocations with Burgers vectors parallel to $[1\bar{1}00]$ or $[0\bar{1}10]$ crystallographic directions. The number of dislocations at the edge increases with the irradiation temperature. More dislocations transfer more material out of the track core, resulting in a larger damaged track region (cf. **Figure 15**).

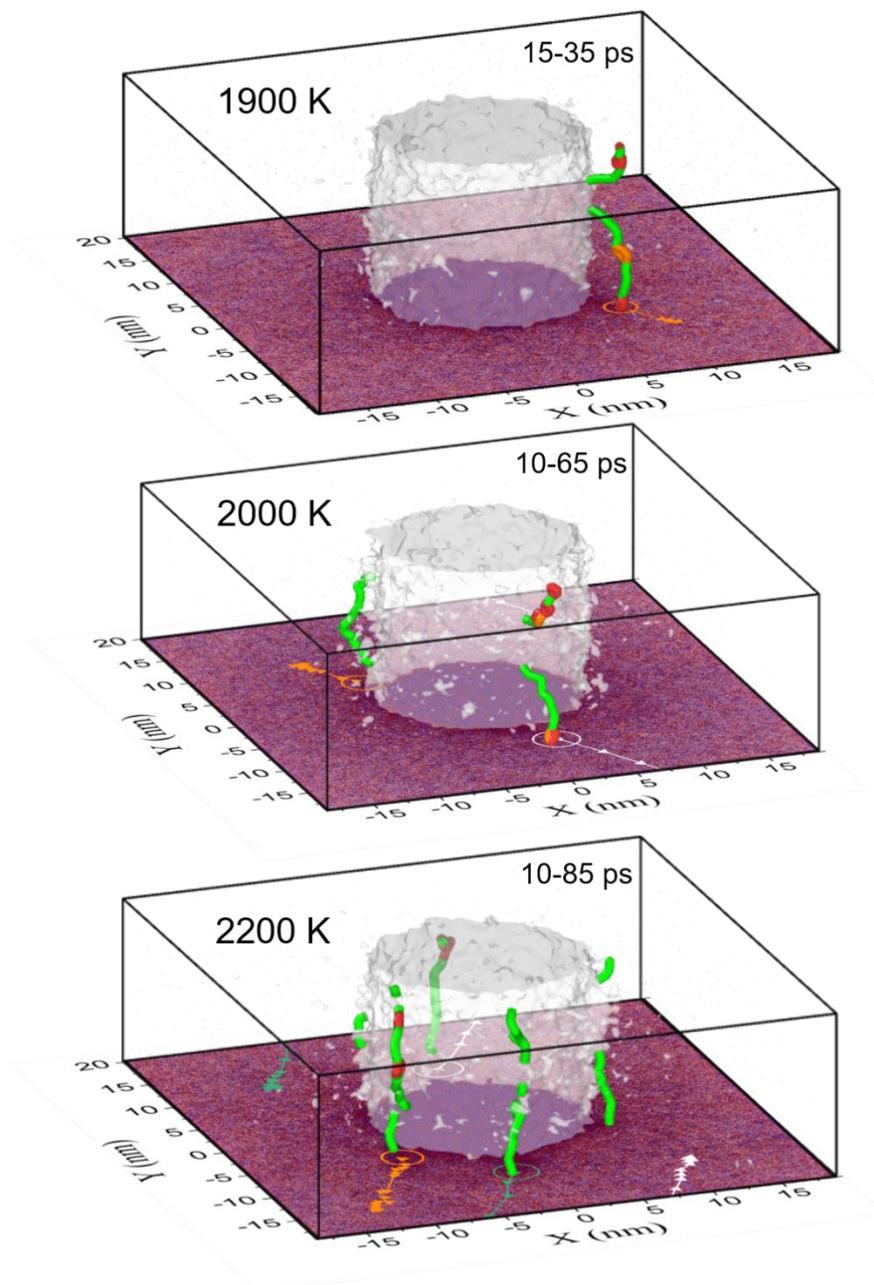

**Figure 17.** *Three-dimensional image of dislocation lines in the SiC supercell after impacts of 710 MeV Bi ion at different irradiation temperatures. The Burgers vector of the green dislocations is directed parallel to the lattice vectors $[0\bar{1}10]$ or $[1\bar{1}00]$, the orange ones parallel to $[\bar{2}110]$ or $[\bar{1}\bar{1}20]$, and the red ones in other directions. A white transparent surface separates the defective region of the crystal around the ion trajectory. Bottom planes show the supercell in the direction perpendicular to the ion incidence. The colored lines in this plane show the trajectories of the dislocations during periods indicated in the figures. Different colors correspond to different trajectories.*

**Figure 18** shows the trajectories of dislocations at different temperatures and simulation times. Note that some dislocations glide away from the track so far that even cross the periodic boundaries of the supercell, which should be taken into account when interpreting the snapshots shown in the figure. In particular, the analysis of the subsequent evolution of the emitted dislocations after the track cooling down to the ambient temperature would require the use of considerably larger simulation cells and is beyond the scope of this work.

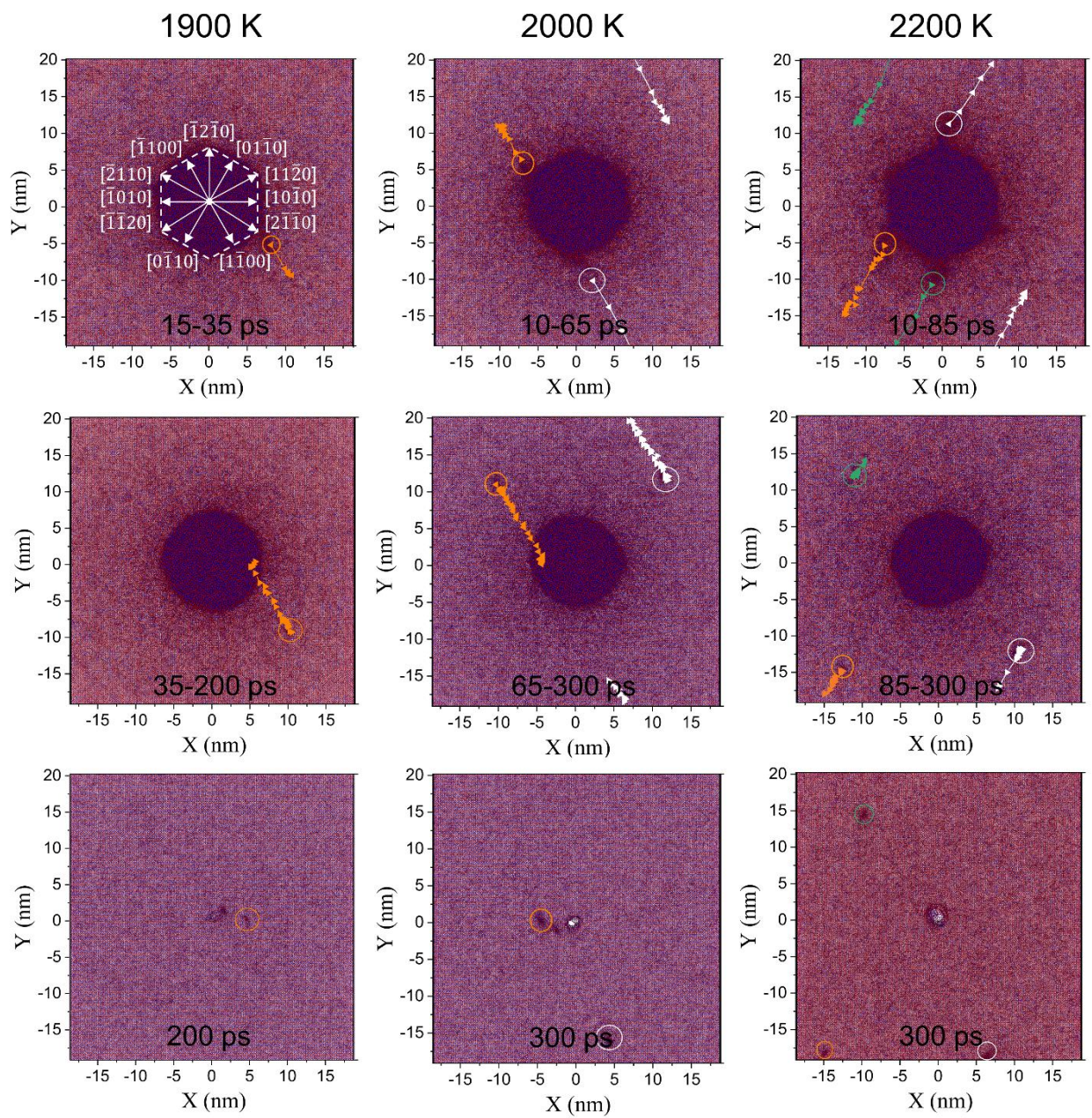

**Figure 18.** *Images of the SiC atomic lattice in the direction of ion incidence after ion impact of 710 MeV Bi ion at temperatures of 1900 K, 2000 K, and 2200 K at different times. The trajectories of lattice distortions near dislocation cores are shown by lines, while triangles indicate the direction of motion at every 5 ps of simulation within the periods marked in the panels.*

# V.    Conclusions

We applied the multiscale TREKIS-3+MD model with the calculated temperature-dependent scattering cross sections to analyze SHI track formation in SiC at different irradiation temperatures. We demonstrated that only Vashishta interatomic potential provides an agreement of MD simulations with the experiments on irradiation of SiC with SHIs.

The density of the energy deposited in the atomic lattice of SiC, as well as the rate of the energy transfer from the excited electronic system in SHI tracks, increase with the irradiation temperature. For 710 MeV Bi ion, this leads to the formation of the track core in SiC at temperatures above ~1800 K, while at lower irradiation temperatures the material is resistant to the irradiation-induced damage.

The damaged track core consists of a chain of cavities with diameters from 0.5 nm to 2 nm surrounded by a cloud of point defects and defect clusters. We have demonstrated that the primary reason for the formation of damaged track cores in SiC at temperatures above the threshold is the mass transport from the melted track region caused by the edge dislocation emission. The result indicates the necessity to introduce temperature criteria of resistance to SHI irradiation of promising materials.

# Code and data availability

The TREKIS-3 [14,15] code, used to model the excitation of the electronic and ionic systems in the SHI track, is available from [69]. The LAMMPS code was used for molecular dynamic calculations [16]. Ovito [70] software was used for visualization of MD calculations, dislocation [71] and defect analysis and surface plotting [72] (with a probe sphere radius $R_\alpha = 3.1$ Å).

# Acknowledgements

This work has been carried out using computing resources of the federal collective usage center Complex for Simulation and Data Processing for Mega-science Facilities at NRC "Kurchatov Institute", http://ckp.nrcki.ru/. NM thanks the financial support from the Czech Ministry of Education, Youth, and Sports (grants No. LTT17015, LM2023068, and No. EF16_013/0001552). The work of R.A. Rymzhanov was funded by the Russian Science Foundation, The Russian Federation (grant No. 23-72-01017, https://rscf.ru/project/23-72-01017/).